\documentclass[11pt,a4paper]{article}

\usepackage{makecell}
\usepackage{comment}
\usepackage[dvipdfmx]{graphicx}
\usepackage{bmpsize}
\usepackage{color}
\usepackage{url}
\usepackage{amssymb}
\usepackage[centertags]{amsmath}
\usepackage{multirow}
\usepackage{subfigure}
\usepackage[affil-it]{authblk}
\usepackage{cite}
\usepackage{notoccite}
\usepackage{lineno}
\usepackage[colorlinks,linkcolor=blue,anchorcolor=blue,citecolor=blue,urlcolor=blue]{hyperref}
\usepackage{bookmark}
\usepackage[T1]{fontenc}
\usepackage[toc,page]{appendix}
\usepackage{float}

\newcommand{\nel}{$\nu_e$}
\newcommand{\nm}{$\nu_\mu$}
\newcommand{\anel}{$\bar{\nu}_e$}
\newcommand{\anm}{$\bar{\nu}_\mu$}

\newcolumntype{C}{>{$\displaystyle}c<{$}}
\newcolumntype{L}{>{$\displaystyle}l<{$}}

\begin{document}

\title{JUNO sensitivity to low energy atmospheric neutrino spectra}
\vspace{0.5cm}
\author{The JUNO Collaboration\footnote{e-mail: juno\_pub\_comm@juno.ihep.ac.cn}}
\date{\vspace{-5ex}}

\maketitle

\begin{abstract}
Atmospheric neutrinos are one of the most relevant natural neutrino sources that can be exploited to infer properties about cosmic rays and neutrino oscillations. The Jiangmen Underground Neutrino Observatory (JUNO) experiment, a 20\,kton liquid scintillator detector with excellent energy resolution is currently under construction in China. JUNO will be able to detect several atmospheric neutrinos per day given the large volume.
A study on the JUNO detection and reconstruction capabilities of atmospheric \nel~and \nm~fluxes is presented in this paper. In this study, a sample of atmospheric neutrino Monte Carlo events has been generated, starting from theoretical models, and then processed by the detector simulation. The excellent timing resolution of the 3'' PMT light detection system of JUNO detector and the much higher light yield for scintillation over Cherenkov allow to measure the time structure of the scintillation light with very high precision.  Since \nel~and \nm~interactions produce a slightly different light pattern, the different time evolution of light allows to discriminate the flavor of primary neutrinos. A probabilistic unfolding method has been used, in order to infer the primary neutrino energy spectrum from the detector experimental observables. The simulated spectrum has been reconstructed between 100\,MeV and 10\,GeV, showing a great potential of the detector in the atmospheric low energy region.
\end{abstract}

\section{Introduction}
\noindent Atmospheric neutrinos are a naturally occurring neutrino source. They originate from the decays of $\pi$ and $K$ produced in extensive air showers initiated by the interaction of cosmic rays with the Earth's atmosphere~\cite{Gaisser2004,HKKM07,GH2002,GaissMod15}. 
The energy spectrum of primary cosmic rays above 100\, MeV can be described by a power law 
$\frac{dN}{dE} \propto E^{-\gamma}$, where the spectral index is $\gamma \simeq 2.7$ for $E\leq 10^6$\,GeV and $\gamma \simeq 3.0$ above that value~\cite{Thebible}. At energies larger than $5 \times 10^8$\,GeV, the spectrum becomes steeper ($\gamma \simeq 3.2$)\cite{Hoerandel_2004gv} and it flattens again $(\gamma \simeq 2.7)$ when $E \geq 3 \times 10^9$\,GeV.
In the interaction of a single high energy Cosmic Ray with the nuclei of the Earth's atmosphere, hundreds or thousands mesons can be produced.  
The atmospheric neutrino energy spectrum spans a wide range from the MeV up to the PeV scale and can be roughly described by a power law~\cite{SK16,IcNuE13,IcNuE15,ICNuMu11,Frejus95}. The spectral index is, in general, steeper than that of primary cosmic rays, since the parent mesons lose a large fraction of their energy before decaying. The spectrum intensity is suppressed at sub--GeV energies reflecting the rigidity cutoff, that describes the shielding provided by the geomagnetic field against the arrival of cosmic rays particles from outside the magnetosphere. Neutrinos originating from muon decays contribute mainly up to a few GeV. The flavor ratio (\nm+\anm)/(\nel+\anel) is around two at $\sim$1\,GeV and increases as the energy increases, since more muons are likely to reach the Earth's surface without decaying. At energies above hundreds of GeV, the decay length of $\pi$ and $K$ becomes longer than their path length in the atmosphere, leading to a neutrino flux reduction. At the highest energies, the decay of heavy charmed mesons is expected to dominate the atmospheric neutrino production. Given the very short lifetime of these particles, the associated neutrino flux is commonly referred as ``prompt''~\cite{promptnu08,promptnu03,prompt01}.

Since the Earth is mostly transparent to neutrinos below the PeV energy scale, an atmospheric neutrino detector is able to see neutrinos coming from all directions. The distance from the production point to the detector varies from $O(10)$ to $O(10^4)$\,km, depending on the zenith angle~\cite{ICOsc18}. The angular distribution has a characteristic shape with an increased flux towards the horizontal direction (with respect to the vertical direction), due to the longer path length of parent particles in the atmosphere. In the sub--GeV energy region there is an asymmetry along the East--West axis, which reflects the azimuthal dependence of the rigidity cutoff of the cosmic rays.

Atmospheric neutrinos were detected for the first time in the 1960s~\cite{AtmoDisc65_1,AtmoDisc65_2}. Further measurements led to the discovery of neutrino flavor oscillations in 1998~\cite{SkOsc98}. Some of the missing pieces in the puzzle of neutrino physics are going to be addressed also by means of atmospheric neutrinos. The field of research is currently very active and several experiments are scheduled in the coming years to answer the unsolved questions. Next-generation detectors for atmospheric neutrino physics plan to significantly improve performances, compared to present ones, by increasing their size and detection granularity. The efforts are mostly concentrated on flavor oscillation physics, pushing the detectors sensitivity for the neutrino mass ordering (MO) and the CP phase $\delta$ in the neutrino sector. The most prominent examples are DUNE~\cite{DUNE_TDR_II}, Hyper-Kamiokande~\cite{HK_DR_18}, INO~\cite{INO17}, ORCA~\cite{ORCA16}, and PINGU~\cite{PINGU17}.

\begin{figure}[t]
\begin{center}
\includegraphics[width=0.49\textwidth]{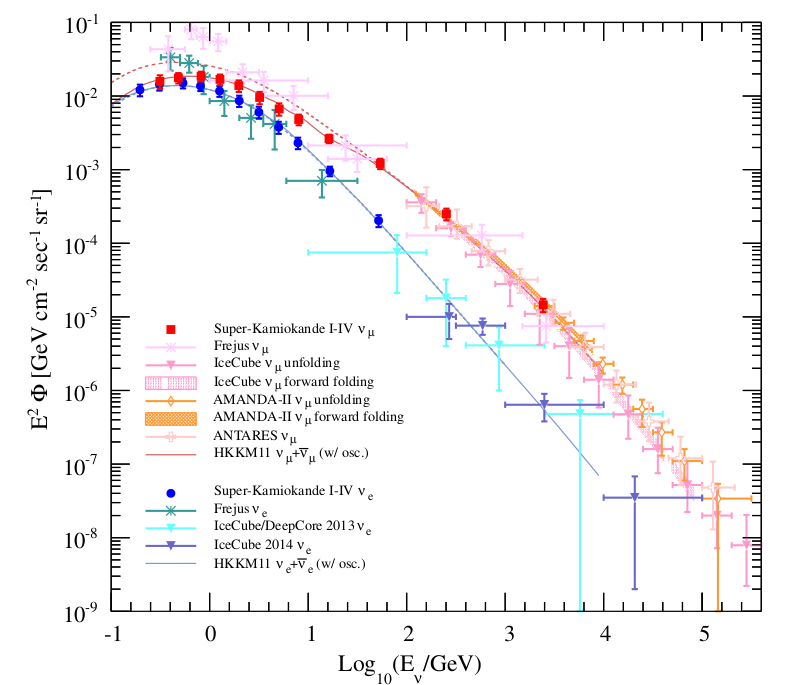}
\caption{Present measurements of the atmospheric neutrino energy spectrum, compared with theoretical predictions. Data and models are reported separately for $\nu_e$ and $\nu_\mu$. Figure from~\cite{SK16}.}
\label{fig:atmo_meas}
\end{center}
\end{figure}

In Figure \ref{fig:atmo_meas}, present measurements of the energy spectrum of atmospheric neutrinos are reported, including predictions from theoretical models.
Measurements performed over the last decades, up to present times, are able to cover a very wide range in the neutrino energy, from several hundreds of MeV to several hundreds of TeV. This sector has been explored predominantly by Cherenkov detectors, such as Super-Kamiokande~\cite{SK16} and IceCube~\cite{IcNuE13,IcNuE15,ICNuMu11,ICNuMu17}.
The Jiangmen Underground Neutrino Observatory (JUNO), currently under construction in China, will be able to detect several atmospheric neutrinos per day. JUNO is going to become the largest liquid scintillator (LS) based detector ever built, having a target LS mass more than one order of magnitude larger than present ones. The large detector mass is one of the key-points for atmospheric neutrino detection, since it is comparable to the largest present water-based detector, Super-Kamiokande. Despite the limited ability of JUNO in tracking single particles after a neutrino interaction, with respect to large Cherenkov detectors, and a slightly reduced accessible statistics, the LS nature of the detector allows more precise measurements towards the low energy region. This sector of the energy spectrum is still not fully covered by present and past experiments. Furthermore, it also corresponds to the region where theoretical models have the largest uncertainties.

The atmospheric neutrino flux measurements by means of the JUNO detector allow to investigate the neutrino MO and the $\theta_{23}$ octant. It is possible to pursue also the CP phase $\delta$ measurement.
In our work, we investigate JUNO's potential for measuring the atmospheric \nel~and \nm~fluxes in the energy range 100\,MeV -- 10\,GeV.

\section{JUNO experiment}
\label{sect:juno}
\begin{figure}[t]
\begin{center}
\includegraphics[width=0.48\textwidth]{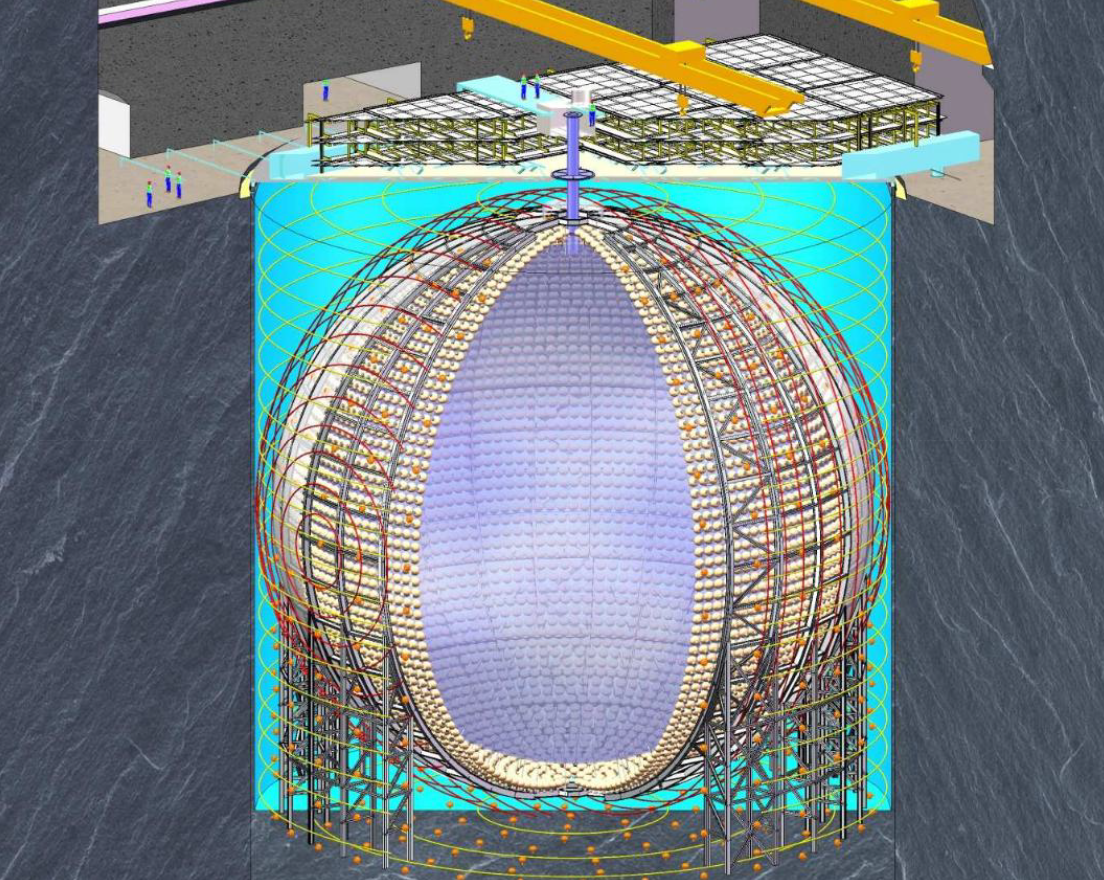}
\caption{Layout of the JUNO detector.}
\label{fig:juno}
\end{center}
\end{figure}

\noindent The JUNO experiment~\cite{YBJuno,JUNO_CDR} is a LS neutrino detector currently under construction in a dedicated underground laboratory (about 700 m deep, 1800 m.w.e.) near Kaiping, Jiangmen city, Guandong province (P. R. China). A sketch of the detector is shown in Figure \ref{fig:juno}. The  central detector (CD) consists of 20\,kton of LS, contained in a 12 cm thick, highly transparent, acrylic sphere with a diameter of 35.4\,m. The light produced in the LS is read out by 17612 20'' high quantum efficiency (QE) photomultiplier tubes (PMTs) and 25600 3'' PMTs, providing a total photo--coverage of more than 75\%. About 13000 of the 20'' PMTs are Microchannel Plate (MCP) PMTs, developed by the JUNO collaboration and currently being produced by the North Night Vision Technology company. The remaining 5000 20'' PMTs consist of the R12860 model produced by Hamamatsu. Both of these PMTs have a 
photon detection efficiency greater than 27\%.
For the 20'' PMTs, the full waveform will be acquired. Their large photon collection area, however, has the consequence of a large dark noise rate, on average of the order of 30\,kHz, and a time resolution on single photo-electrons in the range from 1\,ns to 10\,ns. The additional 3'' PMTs, built by the HZC company, are deployed in the 20'' PMTs' lattice structure, in order to reduce any possible systematics due to the loss of linearity in charge reconstruction and to improve the timing measurements~\cite{SPMTs}. Due to their small area, the 3'' PMTs will operate in digital mode, thus being an independent readout system that can be exploited for cross-calibrating the 20'' PMTs energy response. This feature becomes extremely important for high-energy events, where millions of photons are produced. Furthermore, due to the size difference, the Transit Time Spread (TTS) of 3'' PMTs is of the order of the nanosecond, while the 20'' PMTs one is larger, on average.

The acrylic sphere is surrounded by a stainless steel truss structure, which has a diameter of about 40\,m and constitutes the mechanical support for both the acrylic sphere and the PMTs. The central detector is submerged in a
$\sim$44\,m deep water pool (WP) filled with $\sim$30\,kton of ultrapure water and instrumented with 2400 20'' MCP-PMTs. It acts as an active Cherenkov muon veto and shields the CD against external radioactivity. The walls of the pool are covered with high--reflectivity Tyvek film, in order to increase the photon collection and allowing to veto Cosmic Ray muons with $>$95\% efficiency.  A Top Tracker (TT) is placed on top of the water pool, to improve the total veto efficiency and the reconstruction of atmospheric muons. The TT consists of three layers of scintillator strip detectors, refurbished after the decommissioning of the OPERA experiment Target Tracker~\cite{OperaTT}. It has a granularity of $2.6 \times 2.6$ cm$^2$ and a coverage of about 60\% of the WP top surface.

The JUNO LS mixture consists of three components: Linear Alkylbenzene (LAB) as solvent, 2.5 g/l of 2,5-Diphenyloxazole (PPO) as  scintillation fluor and 3 mg/l of 1,4--Bis(2--methylsyryl) benzene (bis--MSB) as wavelength shifter~\cite{junoscint2020}. This mixture ensures an effective light yield of $\sim10^4$ photons per MeV of deposited energy and an attenuation length greater than 20\,m for 430\,nm photons. The designed radio-purity levels of the JUNO LS are $\mathcal{O}(10^{-16})$ g/g for the bulk $^{238}$U, $^{232}$Th, and $^{40}$K contaminants~\cite{LOMBARDI20196}.
The calibration of the JUNO CD will be performed by four different systems~\cite{junocollaboration2020calibration}. An Automated Calibration Unit (ACU) will deploy different radioactive sources along the detector vertical axis. The ACU system is also designed to deploy a laser source, with a photon intensity that can cover a range from hundreds of keV up to $\mathcal{O}$(TeV) equivalent energy. Two Cable Loop Systems (CLS) will instead place sources across two planes. A guide tube (GT) system, installed on the outer circumference of the sphere, will provide information regarding non-uniformity at the CD boundary. 
A Remote Operated Vehicle (ROV) will finally deploy sources in the whole detector volume. Periodical calibration campaigns will ensure to keep the overall energy resolution around 3\%/$\sqrt{E / \mathrm{MeV}}$ in the MeV energy region, where the analysis for the neutrino MO will focus.
Atmospheric neutrinos interacting inside JUNO can produce different final states, depending on the nature of the interaction they undergo. A first distinction can be done between charged-current (CC) and neutral-current (NC) interactions. In the first case, the lepton of the same flavor of the interacting neutrino is produced and therefore the original neutrino information is preserved. In NC interactions, on the contrary, only a hadronic state is visible and the flavor of the interacting neutrino cannot be inferred. In the energy of interest for atmospheric neutrinos, the dominant interaction is the neutrino-nucleon scattering. The most prominent channels are the elastic and quasi-elastic scattering, the resonant production, and the deep~inelastic scattering~\cite{NeuCS12}. This last classification concerns only about the final hadronic products, which give information about the original energy of the neutrino, but are not sensitive to the interaction flavor. Instead, the CC or the NC nature of the neutrino interaction implies a fundamental difference in the visible products. Apart from the flavor information, the absence of the flavor-corresponding lepton in the final state for NC interactions means also that the neutrino carries away part of its initial energy, which is not released inside the detector. NC events are therefore expected to be concentrated at lower values of the visible energy, while CC ones dominate at higher energy.

\section{Monte Carlo dataset}
\label{sect:simulation}
\begin{figure}[t]
    \centering
    \includegraphics[width=0.51\textwidth]{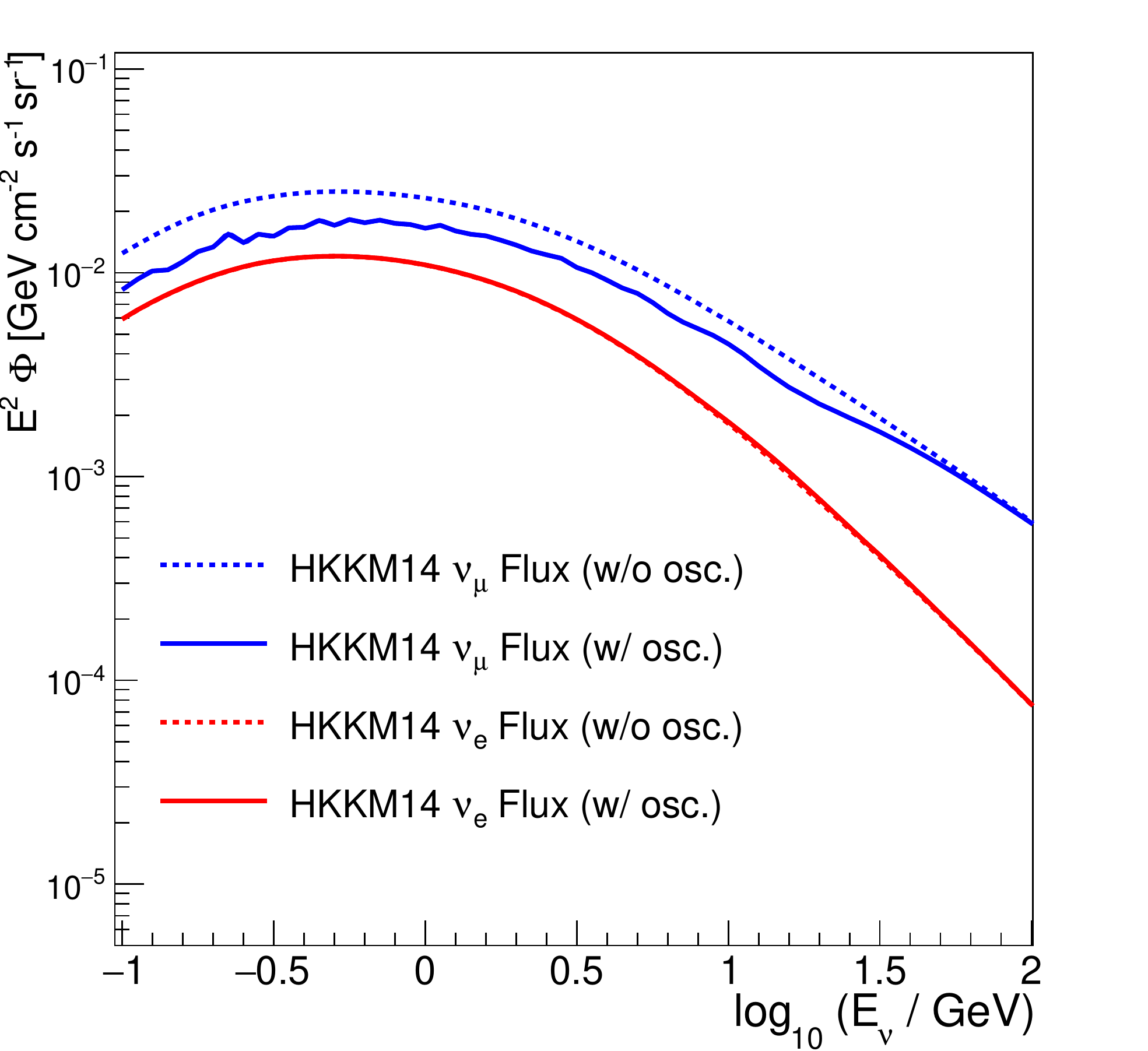}
    \caption{Expected atmospheric (neutrino + antineutrino) flux at the JUNO site, for \nel~(red) and \nm~(blue), according to the HKKM14 model~\cite{HKKM14}. The flux is reported with (full line) and without (dashed line) considering neutrino oscillations.}
    \label{fig:HKKM_flux_JUNO}
\end{figure}

\noindent 
 The study of the atmospheric neutrino flux is usually based on the predictions of the expected flux made by Monte Carlo simulations.  In this work, we consider the predictions from the latest version of the \texttt{HKKM} model~\cite{HKKM14}, which we refer to as HKKM14 hereafter. The model assumes a Cosmic Ray spectrum based on BESS~\cite{Sanuki_2000,HAINO200435} and AMS-01~\cite{AMS-01} measurements. The \texttt{DPMJET-III}~\cite{DPMJET3} and the \texttt{JAM}~\cite{JAM2006} hadronic interaction models are used for the simulation of the interaction with the Earth's atmosphere. The \texttt{HKKM} model provides the calculation of the expected atmospheric neutrino flux at different locations, taking into account the latitude and longitude of the detector. The energy range spans from 100\,MeV up to 10\,TeV. Solar modulation and the asymmetry in the azimuthal distribution are also considered. In the \texttt{HKKM} parametrization, the neutrino flux is calculated at the source and therefore no oscillation effects are included. The HKKM14 atmospheric neutrino flux prediction, calculated at the JUNO experimental site, is shown in Figure \ref{fig:HKKM_flux_JUNO}. Hereafter, \nel~and \nm~will be used to label both neutrinos and antineutrinos of electron and muon flavor, respectively.
In order to get a realistic prediction, neutrino oscillations have been applied to the original flux, including matter effects. The impact of the oscillation effects has been evaluated according to the standard 3-neutrinos mixing scheme~\cite{PDG18}.

The interaction of atmospheric neutrinos with the JUNO detector has been simulated by means of the \texttt{GENIE} Neutrino Monte Carlo Generator~\cite{GENIEpaper,GENIE2MAN15} inside an energy range up to 20\,GeV. The elemental composition of the neutrino target has been set as the one of the JUNO LS (mainly $^{12}$C and $^1$H,
with relative composition of 0.88 and 0.12, respectively). The output of the simulation contains information about the type of interaction that neutrinos undergo, either a CC or a NC one, and the full list of secondary particles and their associated properties (Particle ID, momentum, direction, $\dots$).
The contribution of $\nu_\tau$ CC interactions has been found not to affect the analysis results by independent evaluations and have been therefore not considered in the present work.
Secondary particles produced in the interaction between neutrinos and the JUNO target material have been propagated in the detector by using a \texttt{GEANT4}-based Monte Carlo simulation. The JUNO detector simulation code has been developed within the \texttt{SNiPER} framework~\cite{SNiPER15}. In the detector simulation, several physical processes are included: electromagnetic interaction, decay, hadronic elastic and inelastic interactions, scintillation (including re-emission), Cherenkov emission, and optical absorption.  A detailed optical model, including the optical properties of all the detector materials, is also implemented. 
The output relevant for the analysis includes the timestamp, the number of photoelectrons, and the position of each PMT hit. A data sample of about $5 \cdot 10^5~ {\nu_\mu} + {\nu_e}$ events have been generated, hereafter called large data sample, in order to set up the procedure used to reconstruct the atmospheric neutrino spectrum and to understand the detector response over a large statistics of events.
A sample of 6500 events have been injected in the simulation as a separate Monte Carlo data sample, corresponding to a detector live time of about 5 years. This smaller data sample is hereafter identified as small data sample.

\section{Analysis strategy}
\noindent  As a large LS detector, JUNO achieves its best performance on events which are Fully--Contained within the volume, where a calorimetric measurement can be performed. Partially--Contained events, having some secondaries escaping from the CD active volume, are reconstructed with a worse energy resolution. This analysis therefore targets fully--contained events, to be accounted for reconstruction. This sets an intrinsic upper limit of $\sim$10\,GeV on the \nm~flux, since the high--energy muons produced in a CC interaction always escape the CD volume. For \nel~(\nm), the ``golden'' events consist of \nel~(\nm) fully--contained and CC events and the components to reduce are partially--contained and NC events of all flavor neutrinos. Through--going muons, that may be produced in \nm~interactions with materials surrounding the CD, are not considered in this study.

\subsection{Fiducial cuts} \label{sec:fidcuts}
\noindent Before applying the analysis selection to isolate \nel~and \nm~populations, some preliminary cuts are applied to the large neutrino sample, with the aim of removing low--quality events. A first cut on the interaction vertex position is applied, in order to remove events which release their energy near the edge of the acrylic sphere. These events typically exhibit a loss of linearity between the deposited and the collected energy, because part of the energy is released in the acrylic and water and not in the LS and because the closest PMTs collect a great amount of light and can undergo saturation. A Gaussian smearing with $\sigma$ = 1\,m has been applied to the MC interaction point (hereafter called vertex), in order to reproduce the uncertainty on the reconstructed position. We require that $R_{vertex}$ (i.e. the distance between the vertex and the center of the detector) is less than 16 m to ensure a linear detector response. The precision in the reconstruction vertex at lower energy (in the MeV range) is in general much better than 1\,m; on the contrary, at the GeV energy scale, secondary particles can deposit their energy on a long track and the events can no longer be considered as point-like. It has been checked that even an error of few meters on the vertex position does not affect the performance of the selection procedure. \\ \indent
As described in Section \ref{sect:juno}, the CD is surrounded by a water Cherenkov detector acting as veto for atmospheric muons. Both muons and secondaries coming from partially--contained neutrino events can release a certain amount of energy in the WP and produce a large amount of Cherenkov photons. Therefore, in order to remove partially--contained neutrino events and suppress the atmospheric muon background, we require the total number of hits seen by the water pool veto PMTs $(N_{hits}^{WP})$ to be less than 50, including the contribution from PMT dark noise. This latter term can become important for WP PMTs, because the single-count rate due to WP PMTs dark noise is high (up to several tens of kHz) and the total number of prompt hits from muons Cherenkov light can be small. Hits on WP PMTs are considered in a 200\,ns time window, which is approximately the time needed by a muon to cross the entire detector. The dark hits contribution is simulated on a statistical basis, assuming a binomial distribution.
After applying the fiducial cuts described previously, the simulated large neutrino sample is composed at 97\% of fully--contained event. The remaining partially--contained events are composed at 96\% of $\nu_\mu$ CC interactions. The total efficiency for all \nel~events is 68\%, for \nm~events is 63\%.

\subsection{Atmospheric muon background}
\noindent The atmospheric muon background consists of the secondary muon flux produced after the interaction of cosmic rays with the atmosphere, in the same way as for neutrinos. The JUNO detector location is about 700\,m underground, therefore part of the muon radiation is able to penetrate the rock overburden and release energy inside the detector. The energy released by atmospheric muons inside JUNO is comparable with that of particles coming from atmospheric neutrino interactions (hundreds of MeV - several GeV). Muons can mimic the topology of atmospheric neutrino events and can therefore be a source of background. Although the external water Cherenkov veto is designed to reject these events with high efficiency, the atmospheric muon event rate is several orders of magnitude higher than that from atmospheric neutrino interactions. From preliminary calculations, their event rate inside the JUNO CD is around 3 - 4\,Hz, corresponding to roughly $10^5$ times the atmospheric neutrino event rate, considering that the average energy of atmospheric muons reaching JUNO is 207\,GeV. The desired acceptance rate for the atmospheric muon background must be therefore at least of the order of $10^{-5}$. In order to get a comprehensive picture of the atmospheric muon flux within the framework of this study, a full MC simulation is necessary. Atmospheric muons produce several millions of photons in the JUNO LS and the full detector simulation requires high CPU power and storage. For this purpose, a sample of only $10^5$ muon events has been generated, according to the energy and angular distributions evaluated at the JUNO site. The expected muon flux in the detector is calculated within the JUNO Collaboration according to a parametrized model at Earth surface~\cite{GaisserMod15} and simulating muons propagation through matter~\cite{MUSIC09}. A detector simulation has been performed. Atmospheric muons in JUNO appear as high-energy tracks which release a large amount of energy both in the WP and in the CD. The fiducial cuts described in Section \ref{sec:fidcuts} require instead a low collected light inside the WP.

Hereafter, the readout charge of the event, in terms of the number of PEs collected by CD 20'' PMTs, is called NPE. NPE represents the observable used to reconstruct the neutrino energy. The same fiducial cuts have been applied to the muon sample, with the additional request of more than $10^5$ NPE, which is the region of interest for the analysis. An acceptance of $< 2.3 \cdot 10^{-5}$ at 90\% confidence level is achieved. The accuracy in the estimation of the acceptance will be improved by increasing the Monte Carlo statistics.

\subsection{Neutrino flavor identification} \label{sec:flav_id}
\noindent As mentioned above, \nel~(\nm) CC interactions are the preferred detection channels, since the corresponding charged leptons have very different behaviours. Electrons lose energy quickly via bremsst\-rahlung and ionization and even at GeV energies their track length is no more than 1-2 meters. On the contrary, muons with energy greater than 1\,GeV have longer tracks inside the detector volume. Low--energy muons, moreover, may decay inside the scintillator volume and give a delayed energy release from the Michel electron. The above differences make \nm~CC events more extended in time and space, with respect to \nel~CC events. The latter component has indeed a much shorter evolution. Hadronic particles are common to all classes of events and make up the visible part of NC events. Hadrons, in general, have a long energy release, because of their interactions and decays.

\begin{figure*}[t]
\begin{center}
  \includegraphics[width=0.85\textwidth]{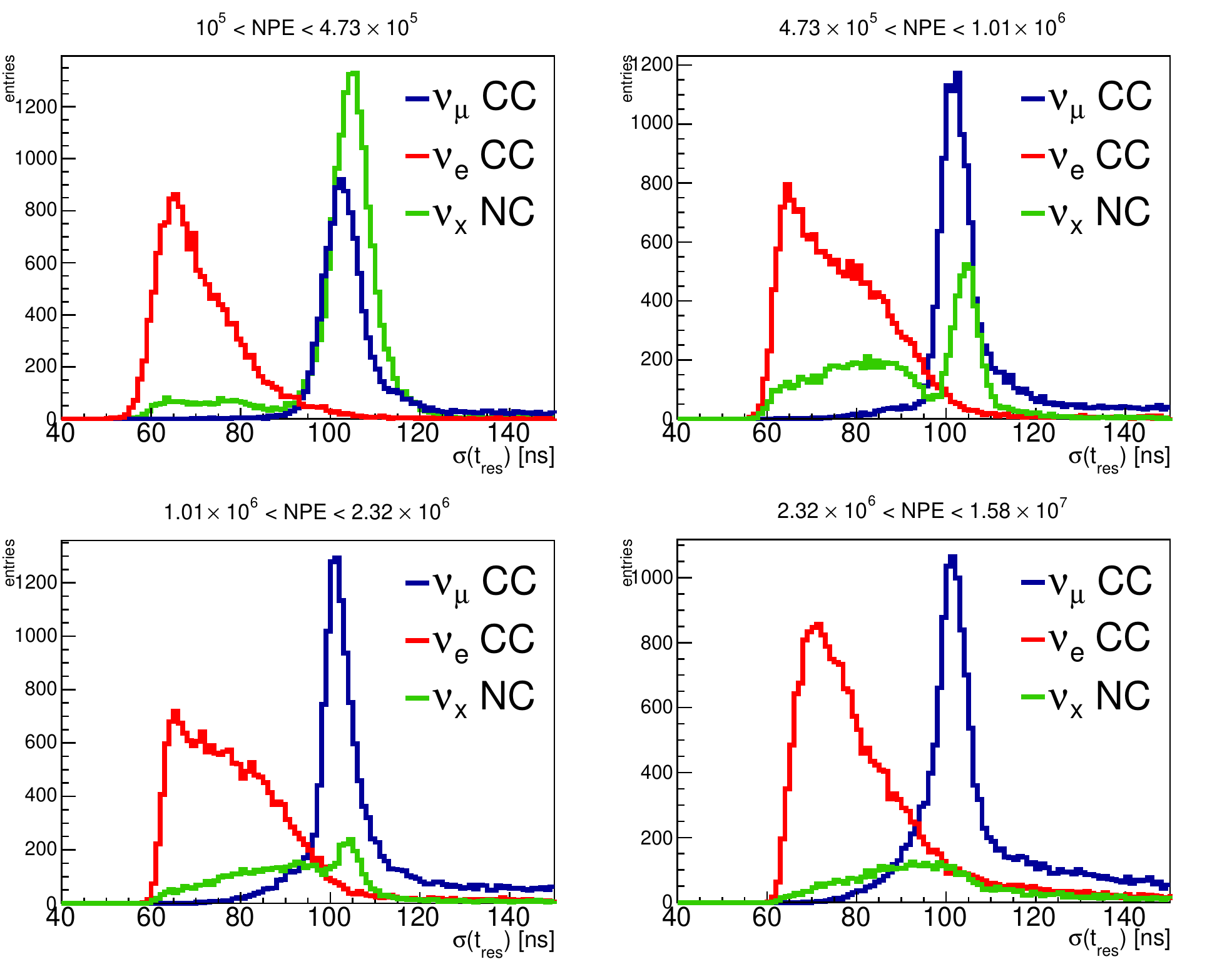}
  \caption{Distribution of $\sigma(t_{res})$ for \nm~CC (blue), \nel~CC (red) and NC events (green), for different ranges of NPE.}
  \label{fig:tres}
\end{center}
\end{figure*}

The event time profile can be therefore exploited to discriminate between different classes of events~\cite{procRICAP}. A high--precision measurement of the photon arrival time is an important requirement. For this reason, the timing information is taken from the data of the 3'' PMT system of the JUNO detector, which have a low TTS value. A Gaussian smearing with a typical width $\sigma = 1.6$\,ns (taken from preliminary measurements) has been applied to the true Monte Carlo hit time over each 3'' PMT. In order to be aligned to a realistic DAQ window, only events inside a 1.2\,$\mu s$ time window have been considered. A time residual $t_{res}$ is then defined for each hit on the i--th 3'' PMT as:
\begin{equation}
    t_{res}^i = t_{hit}^i - \Big(\frac{n\cdot R_V^i}{c}\Big),
\end{equation}
where $t_{hit}^i$ is the hit time on the i--th 3'' PMT, $n$ is the refraction index of the JUNO liquid scintillator and $R^i_V$ is the distance between the reconstructed  vertex position and the i--th 3'' PMT. The time profile of the scintillation light emitted by \nel~and \nm~CC events is different and the latter has a more prominent tail; therefore, the RMS of the $t_{res}$ distribution - hereafter called $\sigma(t_{res})$ - over the fired 3'' PMTs can be used as discrimination parameter. In Figure \ref{fig:tres}, the $\sigma(t_{res})$ distribution is reported for the three populations: $\nu_\mu$ CC, $\nu_e$ CC, and NC events. The variable is also reported separately in 4 different intervals of NPE, selected such as to have equal statistics in each of them.
The plots in Figure \ref{fig:tres} show a good separation between the \nel~CC and the \nm~CC component, over the whole energy range. The NC component appears to be overlapped mainly to \nm~CC events, with a tail also in the \nel~CC region. The reason is that a large fraction of the hadronic component of the secondaries is made of pions, that either decay to \nm~+ $\mu$ ($\pi^+ / \pi^-$) or to two gammas ($\pi^0$). The first category is almost indistinguishable from \nm~CC events, while the second one results in electromagnetic showers and resembles the \nel~CC component. The relative weight of charged and neutral pions in the final state changes across the energy, as well as that of nucleons. This feature is at the origin of the different shape of the $\sigma(t_{res})$ distribution for NC events in Figure \ref{fig:tres}, since each of the four bins of NPE corresponds to a different energy interval. Protons and neutrons, moreover, have a time profile similar to the one of muons, because they result in a long-lasting energy release inside the LS. The contribution of NC component, however, becomes less significant at high energy, due to its steeper spectral shape. In NC events, indeed, part of the initial energy of the interacting neutrino is carried away by the neutrino itself and is not deposited inside the detector.
Given the different features, two separate selection criteria are used to maximize CC events. In order to separate $\nu_e$ events, a value of $\sigma(t_{res})$ < 75\,ns is required. The cut results in an efficiency for $\nu_e$ events $\simeq$42\%, with respect to the large sample after fiducial cuts, and a residual contamination from $\nu_\mu$ less than 6\%. A requirement of $\sigma(t_{res})$ > 95\,ns is required to isolate \nm~events. In order to reduce the contribution from NC events at low energy, an additional requirement of NPE $\geq 5\times10^5$ has been set for \nm~selection, thus limiting the analysis to events with a neutrino energy $\gtrsim$ 400\,MeV. The efficiency for $\nu_\mu$ events is 85\% with respect to the large sample after fiducial cuts and the residual $\nu_e$ contamination is less than 20\%. The residual NC events are populated both by \nel~and \nm.

The flavor identification procedure has been extensively checked by means of an independent analysis. A variation of the $\sigma(t_{res})$ cut has been applied and the resulting efficiency and contamination are in a very good agreement within the statistical and systematic errors reported in this work.

\begin{figure}[ht]
    \centering
    \includegraphics[width = 0.495\textwidth]{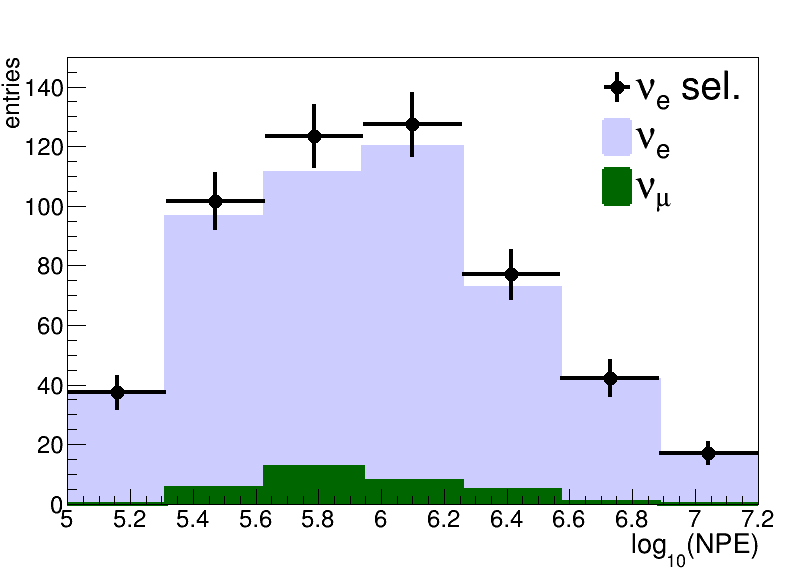}
    \includegraphics[width = 0.495\textwidth]{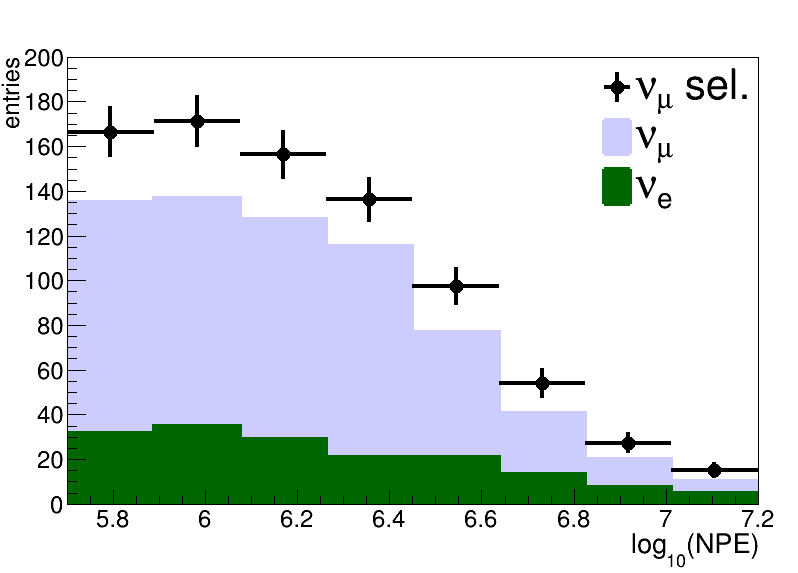}
    \caption{Distribution of the observable $\log_{10}$(NPE) in the analysis bins, after the $\nu_e$ selection (left) and the $\nu_\mu$ selection (right). The black dots represent the number of selected events in every bin $j$, with associated statistical error. The filled histograms reproduce the bin composition, in terms of the correct flavor (light blue) and wrong flavor (green).}
    \label{fig:npespec}
\end{figure}

In order to test the JUNO performance in reconstructing the atmospheric neutrino flux, we used the small Monte Carlo sample corresponding to $\sim$5 years of data-taking described in Section \ref{sect:simulation}.
The energy range considered for the atmospheric \nel~flux is $[-1.00,1.05]$, expressed in $\log_{10}(E_\nu/\mathrm{GeV})$ units and is divided in seven bins. The corresponding $\log_{10}$(NPE) range is $[5.0,7.2]$, divided in seven bins as well. Similarly, the energy range for \nm~is $[-0.30,1.05]$, divided in seven bins, and the corresponding $\log_{10}$(NPE) range is $[5.7,7.2]$, divided in eight bins. The distribution of $\log_{10}$(NPE) is reported in Figure \ref{fig:npespec}, in the bins used in the analysis. 
A summary of the small sample population is in Table \ref{tab:seltable}, as a function of the flavor and of the cuts applied, in the NPE regions considered.

\begin{table}[h!]
  \centering
  \caption{Summary of selections flow for $\nu_e$ and $\nu_\mu$ fluxes, in terms of number of events in the analysis region, applied to the small data sample corresponding to $\sim$5 years of data taking. The values are reported before the selections, after the fiducial cuts and after the $\sigma(t_{res})$ selection. The residual background is also reported.}
\begin{tabular}{c | c c}
&  ${\nu_e}$&  ${\nu_\mu}$ \\
\hline
\rule{0pt}{3ex}
Events injected in the simulation& \multicolumn{2}{c}{6500}\\
\hline
Charge region&  1725&   1241\\
Fiducial cuts&  1167&   773\\ 
$\sigma(t_{res})$ cut&  495&    661\\
\hline
Residual background&   30& 163\\
  \hline
\end{tabular}
\label{tab:seltable}
\end{table}

\subsection{Unfolding}
\label{sect:unfolding}
\noindent The determination of the atmospheric neutrino energy spectrum, starting from the detector experimental observables, is a classical unfolding problem. In this case, the true spectrum is deconvolved from the distribution of the experimental observables, knowing the detector response. In the classical fitting method, on the other hand, the true distribution is extracted from the observables by directly comparing the experimental distribution with the results of a model prediction. The main benefit of the unfolding is that it does not require a particular choice of the spectrum parametrization. In a liquid scintillator detector like JUNO, the main observable for the energy reconstruction is the total number of photoelectrons NPE detected by the 20'' PMTs. This value is related to the total energy deposit in the LS and therefore to the neutrino energy. The neutrino energy spectrum $E_{\nu}$ is then unfolded from the NPE spectrum. In general, the observable NPE spectrum $N$ can be expressed in terms of the primary neutrino spectrum $E$ as
\begin{equation}
\label{eq:response}
    N_j = \sum_i A_{ji}E_i,
\end{equation}
where $A_{ji}$ is the likelihood matrix, which can be estimated by means of a full detector simulation. The relationship in Eq. \ref{eq:response} can be inverted by using the unfolding matrix $U_{ij}$:
\begin{equation}
    \label{eq:unfolding}
    E_i = \sum_j U_{ij}N_j.
\end{equation}
The unfolding matrix $U_{ij}$ can be evaluated by means of an iterative Bayesian procedure~\cite{DAGOSTINI}. In this case, the likelihood matrix $A_{ji}$ can be expressed as the probability $P(N_j|E_i)$ of detecting an event in the $j$--th bin of the NPE spectrum $N_j$ produced by the interaction of a neutrino in the $i$--th bin of the energy spectrum $E_i$. The values of $P(N_j|E_i)$ are evaluated by means of a Monte Carlo detector simulation and normalized as $\sum_j A_{ji} = 1-\varepsilon_i$, where $\varepsilon_i$ takes into account the inefficiency 
in measuring the energy $E_i$.
 The wrong--flavor events are also included in $A_{ji}$. Using Bayes' theorem, the unfolding matrix $U_{ij}$ can be written as:
\begin{equation}
    U_{ij} = P(E_i|N_j) = \frac{P(N_j|E_i)P_0(E_i)}{\sum_iP(N_j|E_i)P_0(E_i)}.
\end{equation}
The prior $P_0(E_i)$ is the probability for a single event to fall into the $i$--th energy bin. Once the unfolding matrix is known, a \emph{first} estimation of the spectrum can be produced:
\begin{equation}
    \hat{E}_i = \sum_j P(E_i|N_j)N_j.
\end{equation}
The normalized values of $\hat{E}_i$ are  used iteratively as the new set of probabilities $P(E_i)$, in order to obtain an updated value of $P(E_i|N_j)$ and therefore of $\hat{E}_i$. The particular choice of the prior and the number of iterations may cause a small bias on the shape of the unfolded spectrum. A small number of iteration may not reflect the information given by the data, while a high number of iteration may amplify statistical fluctuations and distort the spectrum. Since the Bayesian method is strongly data driven, the effect of the particular choice of the prior is in general small, but is still taken into account as a source of systematic uncertainty. The prior should reflect, in principle, the best knowledge of the primary spectrum. The minimum bias is then achieved by adopting the true MC distribution. The strong data--driven nature of the iterative Bayesian method ensures very good results after few iterations. In this work, two iterations have been
performed. A soft smoothing is applied to the first value of the probability $P_1(E_i)$. As prior
distribution, the HKKM14 model has been used.
Further details are given in Section \ref{sect:err}.

\begin{figure}[t]
    \centering
    \includegraphics[width = 0.495\textwidth]{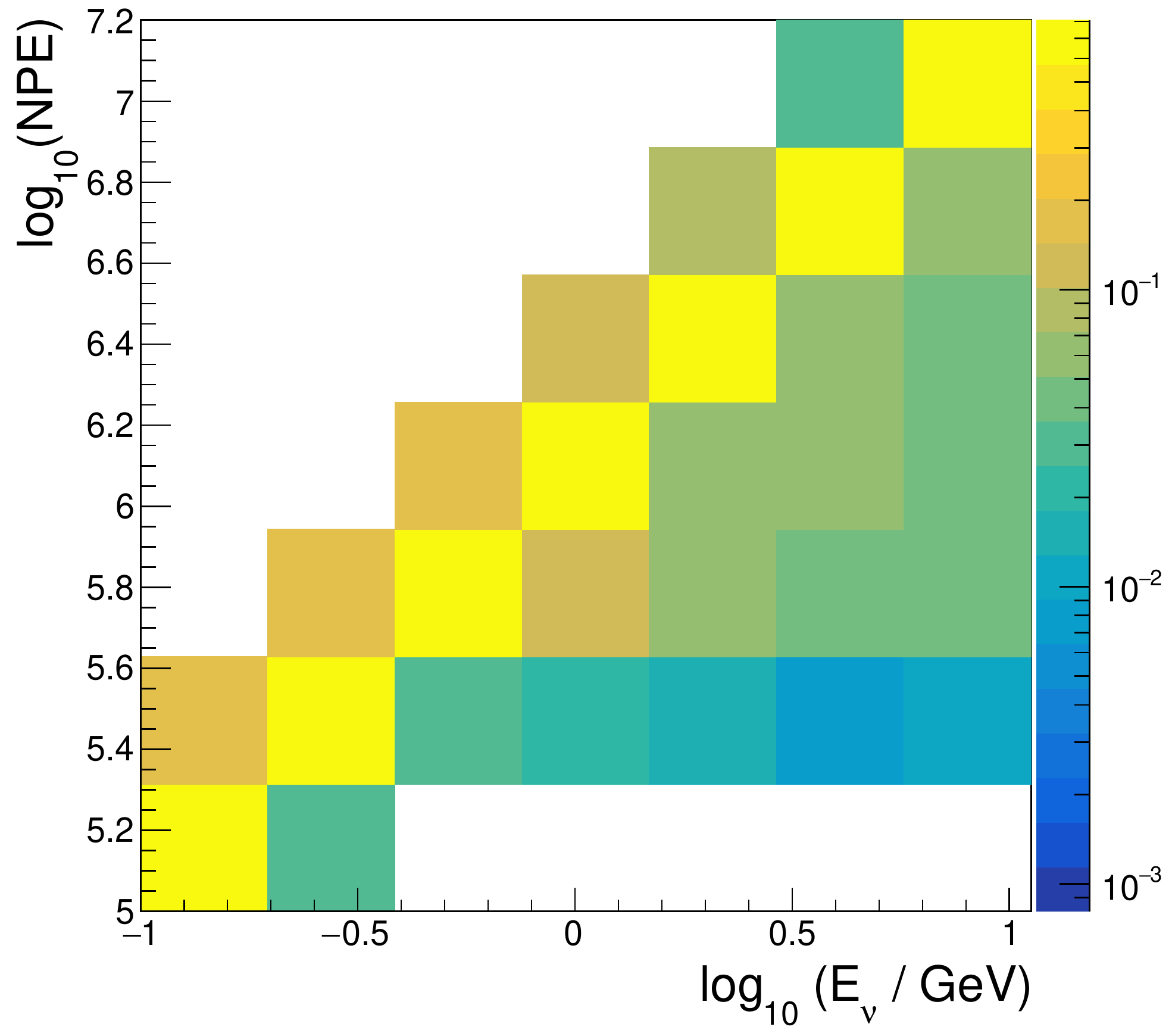}
    \includegraphics[width = 0.495\textwidth]{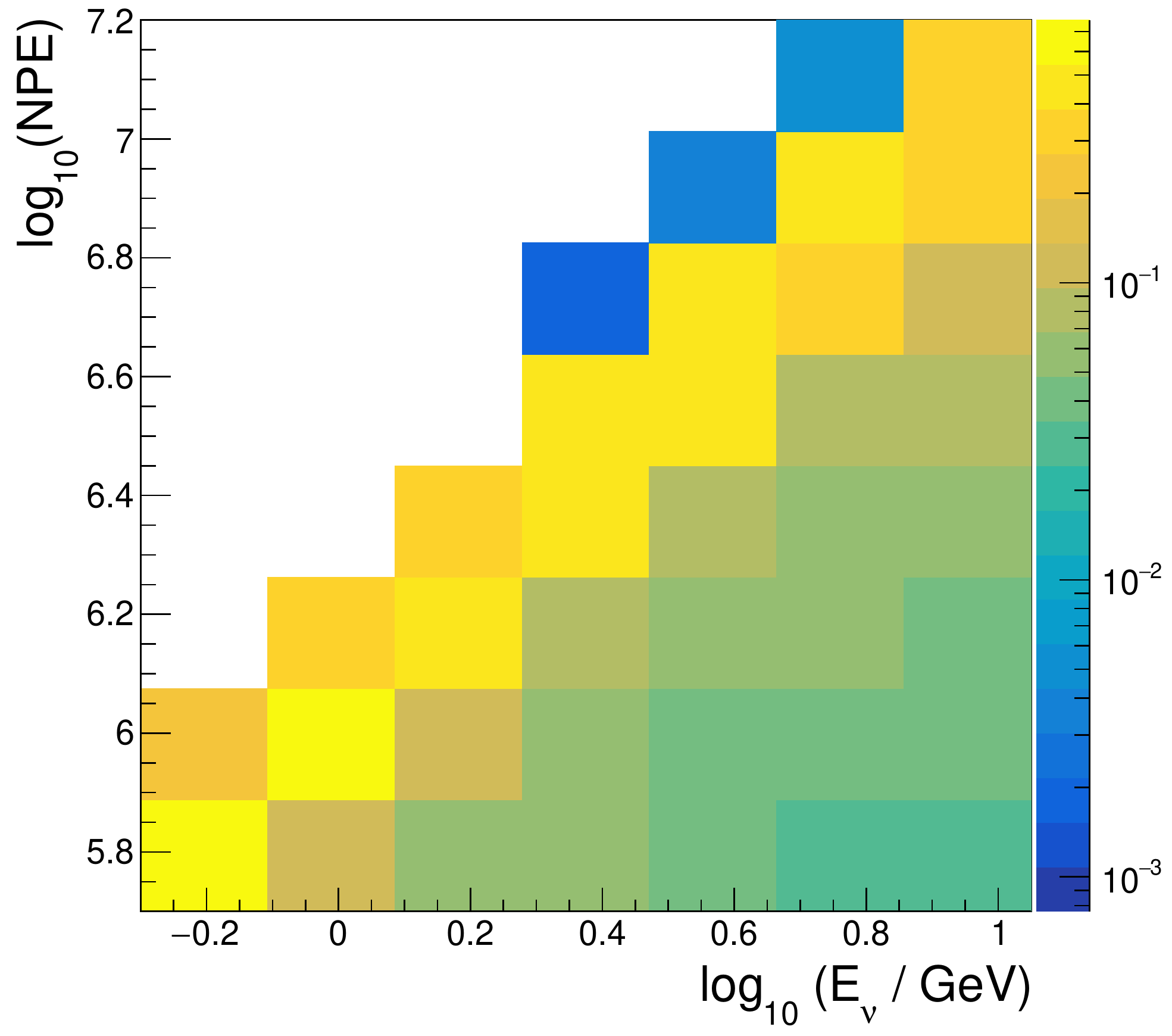}
    \caption{Likelihood matrix for \nel~(left) and \nm~events (right).}
    \label{fig:likelihood}
\end{figure}

 Figure \ref{fig:likelihood} shows the likelihood matrix for both \nel~and \nm~events, evaluated according to the binning described in Section \ref{sec:flav_id} and including the contribution of the background.

\subsection{Uncertainties}
\label{sect:err}
\noindent The total uncertainty on the atmospheric neutrino spectrum reconstruction is evaluated in each energy bin, including both contributions from statistics and systematic effects.
\paragraph{\textbf{Statistics}}
The statistical uncertainty is due to the stochastic fluctuations that occur in the data bins. The amount of this fluctuations is visible in Figure \ref{fig:npespec}, for each observable bin. In order to evaluate their impact in the final unfolded spectrum, 1000 toy data sets have been generated, each time varying the bin content according to a Poisson distribution. 
The final distribution in each bin of the unfolded spectrum is then fit with a Gaussian function, whose $\sigma$ is quoted as the statistical uncertainty. The statistical contribution ranges from 5\% in the bins with highest statistics up to $\sim$15\% in the highest-energy bins.
\paragraph{\textbf{Selection criteria}}
The selection procedure is in general not intended to produce any bias on the final sample. As explained in Section \ref{sec:fidcuts}, fiducial cuts have been
used in the unfolding procedure in order to improve the accuracy of the 
probability evaluations. The energy range of the final reconstructed 
spectrum is well contained inside the energy range of the 
Monte Carlo generated events, guaranteeing  that the fiducial cuts do not 
introduce any bias. The neutrino flavor identification based on the time residual selection, on the other hand, could bring some uncertainty in the data bins where the statistics is low: an even small variation in the chosen cut value of $\sigma(t_{res})$ could result in a substantially different value of the unfolded flux, due to the wide stochastic fluctuations. The whole analysis has been therefore performed by varying the nominal cut value of $\sigma(t_{res})$ by 1\,ns steps in a [\,-5\,ns,~+5\,ns\,] time window. The differences in the unfolded flux are relevant in the bins with less statistics, for the reasons explained above. The total contribution to the bin uncertainty is evaluated as the standard deviation of the flux values distribution in each bin.
\paragraph{\textbf{Flavor oscillation}}
The current uncertainties on the global fit oscillation parameters are reported in Table \ref{tab:nuparams}, which are assumed to be Gaussian. A toy MC has been used to generate 1000 data sets, randomly varying the oscillation parameters within the experimental uncertainties, including the mass ordering and assuming no correlation. The final distribution in the unfolded flux is fit in each bin with a Gaussian function. Since the resulting dispersion is rather small in every bin, the total per-bin uncertainty contribution is quoted as the displacement of the distribution fit peak with respect to the nominal flux value. The total contribution from oscillation parameters uncertainty is estimated to be below 1\% on the entire spectrum. The only exception is the first bin of the $\nu_\mu$ spectrum, where oscillation effects are not negligible, which results into an uncertainty corresponding to a $\sigma$ of 1.2\%.
\begin{table}[t]
  \centering
  \caption{Best-fit neutrino oscillation parameters with associated $3\sigma$ experimental uncertainty. Results are given both assuming normal ordering and inverted ordering. Table is taken from~\cite{PDG18}.}
\begin{tabular}{ l c c }
  Parameter & NH & IH\\
\hline
\noalign{\vskip 0.5mm}
  $\sin^2\theta_{12}$ & \multicolumn{2}{c}{$0.297^{+0.057}_{-0.047}$}\\[2mm]
  $\sin^2\theta_{23}$ & $0.425^{+0.19}_{-0.044}$ & $0.589^{+0.047}_{-0.205}$ \\[2mm]
  $\sin^2\theta_{13}$ & $0.0215^{+0.0025}_{-0.0025}$ & $0.0216^{+0.0026}_{-0.0026}$ \\[2mm]
  $\delta~/{\pi}$ & $1.38^{+0.52}_{-0.38}$ $(2\sigma)$ & $1.31^{+0.57}_{-0.39}$ $(2\sigma)$ \\[2mm]
  $\Delta m^2_{21}~/10^{-5}\,\mathrm{eV}^2$ & \multicolumn{2}{c}{$7.37^{+0.59}_{-0.44}$} \\[2mm]
  $\Delta m^2_{32}~/\,10^{-3}\mathrm{eV}^2$ & $2.56^{+0.13}_{-0.12}$ & $2.54^{+0.12}_{-0.12}$ \\[2mm]
  \hline
\end{tabular}
\label{tab:nuparams}
\end{table}
\paragraph{\textbf{Cross--section}}
The uncertainties on neutrino cross-section impact directly on the number of observed events. In the MC simulation process, as described in Section \ref{sect:simulation}, neutrino interactions are managed by the \texttt{GENIE} software. The full list of uncertainty sources considered by \texttt{GENIE} is provided in~\cite{GENIE2MAN15}. A comprehensive handling of the whole list is not trivial, since it requires the simultaneous calculation of modified interaction probabilities in a wide parameter space. In this study, the evaluation of the cross-section uncertainty is based on experimental measurements provided by the T2K Collaboration~\cite{T2KNueXSec14,T2KNuMuCCQEXSec15,T2KNuMuCC1PiXSec17}, extrapolated from the associated data releases. Assuming the uncertainty on the measured cross section values to be Gaussian, the related visible spectrum is modified accordingly, within 1$\sigma$ interval. The propagated uncertainty on the unfolded flux is evaluated by unfolding 1000 toy MC data sets, with NPE bin contents altered according to random variations of the cross section parameters. The unfolded spectra distributions are then fit in each bin with a Gaussian function, whose $\sigma$ is quoted as the related uncertainty contribution. The uncertainty in the neutrino cross-section values has a large impact in the final reconstructed flux, up to 20\%.
\paragraph{\textbf{Unfolding procedure}}
Although the iterative Bayesian unfolding method is data-driven, the particular MC sample may have an influence on the final result. This means that the initial estimation of the likelihood matrix may have an intrinsic bias, as well as the choice of the prior. The relative impact should be small, but it can have an impact in the unfolding bins with low statistics. 
The net effect cannot be exactly computed, but a reliable estimation can be 
achieved by unfolding modified data sets, generated by assuming a primary MC distribution reasonably far from the one used to evaluate the probabilities.

The modified spectra are produced from the original MC by means of a re-weighting procedure. The new spectrum can be expressed in the $i$-th unfolding bin as:
\begin{equation}
\label{eq:fluxvar}
\Phi_{\nu_i}^{\rm MOD}=(1+\alpha)\left(\frac{E_{\nu_i}}{1\,{\rm GeV}}\right)^{\gamma} \Phi^{\rm MC}_{\nu_i},
\end{equation}

where $\alpha$ acts on the absolute normalization and $\gamma$ on the shape of the primary spectrum. These two parameters are considered to range in the following intervals: $\pm$0.05 for $\alpha$ and $\pm$0.2 for $\gamma$.

\begin{figure}[t]
\begin{center}
\includegraphics[width=0.495\textwidth]{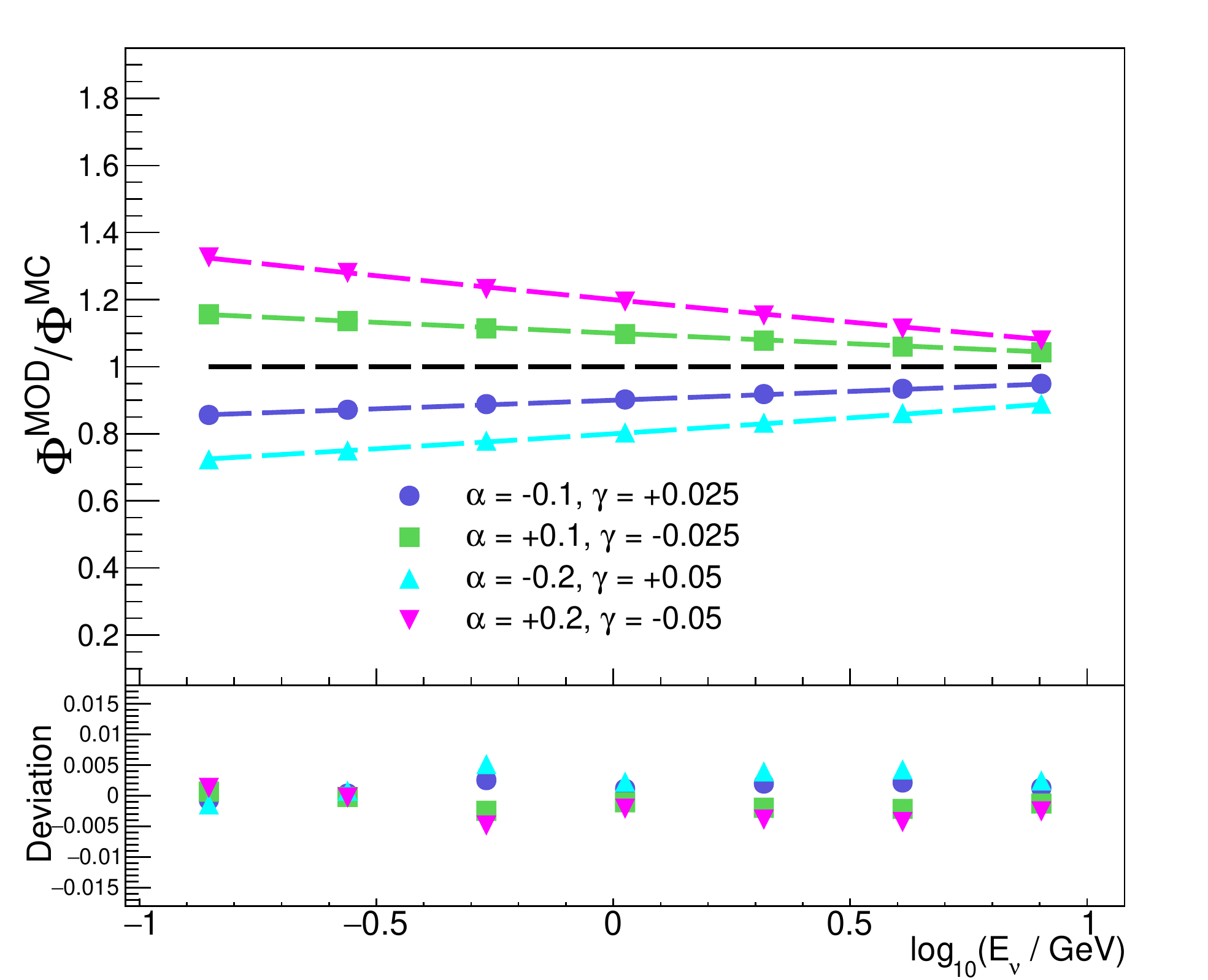}
\includegraphics[width=0.495\textwidth]{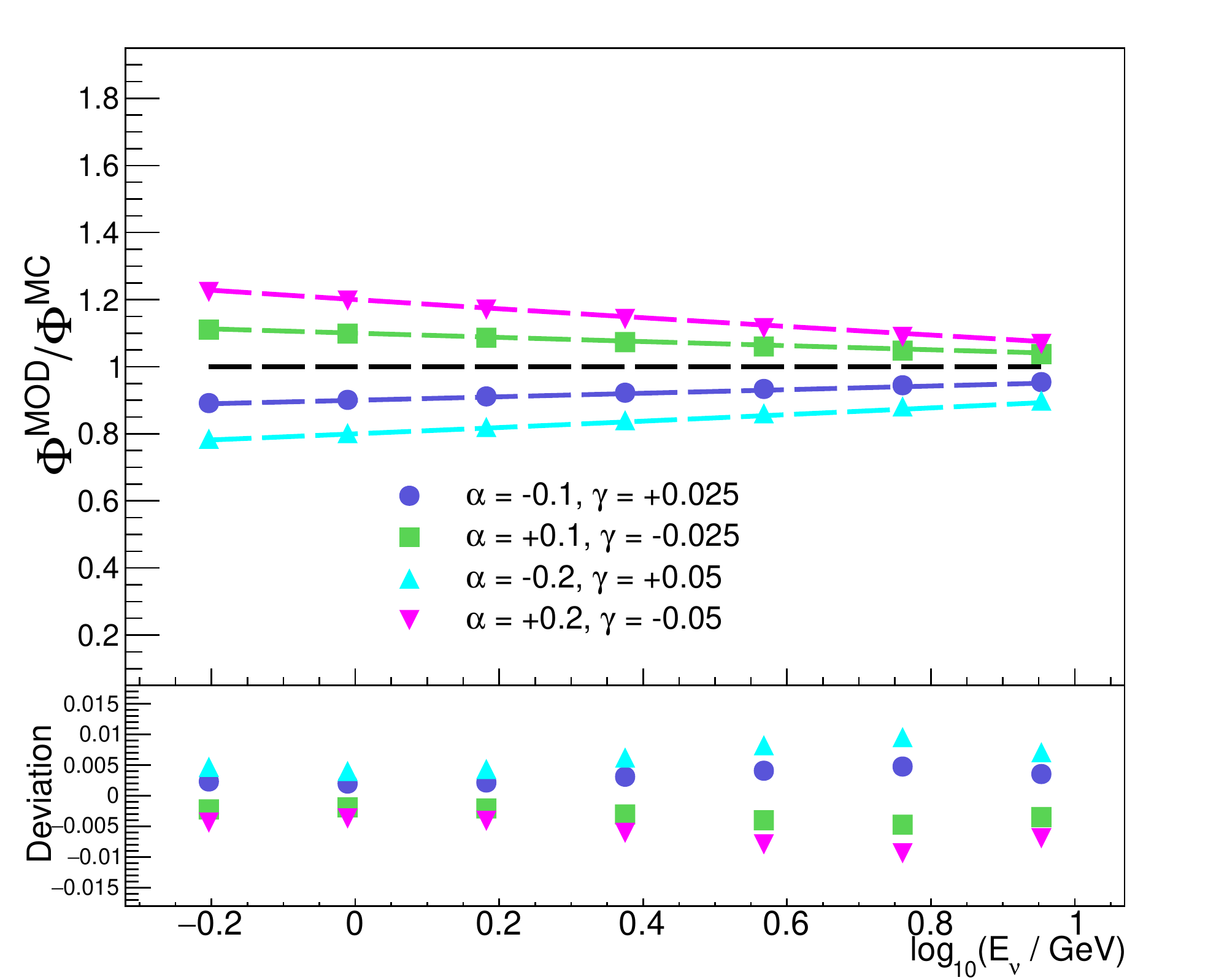}
\caption{Top panels: modified unfolded spectra (markers), together with the corresponding input (dashed lines).  Bottom panels: relative deviation. Left: $\nu_e$ spectra; right: $\nu_\mu$ spectra. $\Phi^{MC}$ represents the nominal flux model~\cite{HKKM14}, which is reported as the black dashed line. Four sets of modified spectra are plotted, whose $\alpha$ and $\gamma$ values are reported in the figures.}
\label{fig:systspec}
\end{center}
\end{figure}
The size of variation corresponds approximately to a 1$\sigma$ 
uncertainty interval in the predicted spectra. In Figure \ref{fig:systspec} 
the comparison between each toy data sample and the corresponding 
unfolding result is reported, together with the fractional deviation between
the input and the unfolded result. The deviation is below 1\% and turns out to be slightly higher in the case of maximum variation of $\alpha$ and $\gamma$ in the bins with lower statistics. The conditional probabilities used in the unfolding 
procedure have been carefully evaluated by using different methods for both \nel~and \nm~samples. The relative deviations obtained for different energy bins and different NPE bins are reported in Table \ref{tab:etable}, as an example, for the \nel~sample. The effect on the obtained spectra turns out to be negligible. 

\begin{table}[t]
  \centering
    \caption{Relative deviations of the conditional probabilities evaluated by using the \nel~sample.}
    \scalebox{0.87}{
  \begin{tabular}{l|c c c}\hline
  \diaghead{\theadfont Diag ColumnmnHead II}
  {$\log_{10}$(NPE)}{$\log_{10} (E_\nu$ / GeV)}
  &  -1 -- -0.5&   -0.5 -- 0.36& 0.36 -- 1.1\\
  \hline
  5 -- 5.5&  0.030 &   0.055 &    0.040 \\
  5.5 -- 6.3& --  &   0.005 &   0.075\\ 
  6.3 -- 7.2&  -- &   --    &   0.005\\
  \hline
  \end{tabular}
  }
  \label{tab:etable}
\end{table}
\par
\begin{figure}[t]
\begin{center}
\includegraphics[width=0.495\textwidth]{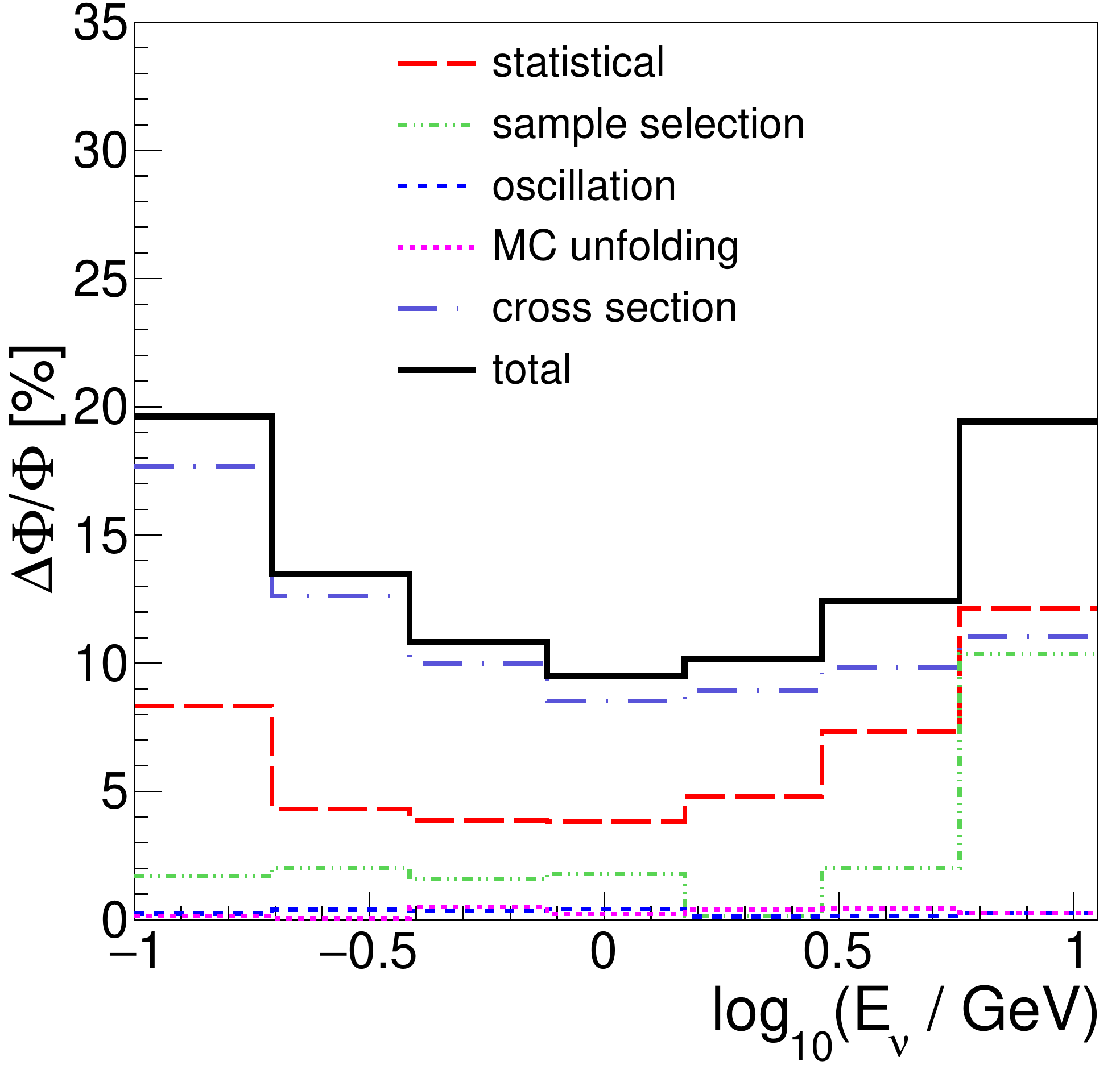}
\includegraphics[width=0.495\textwidth]{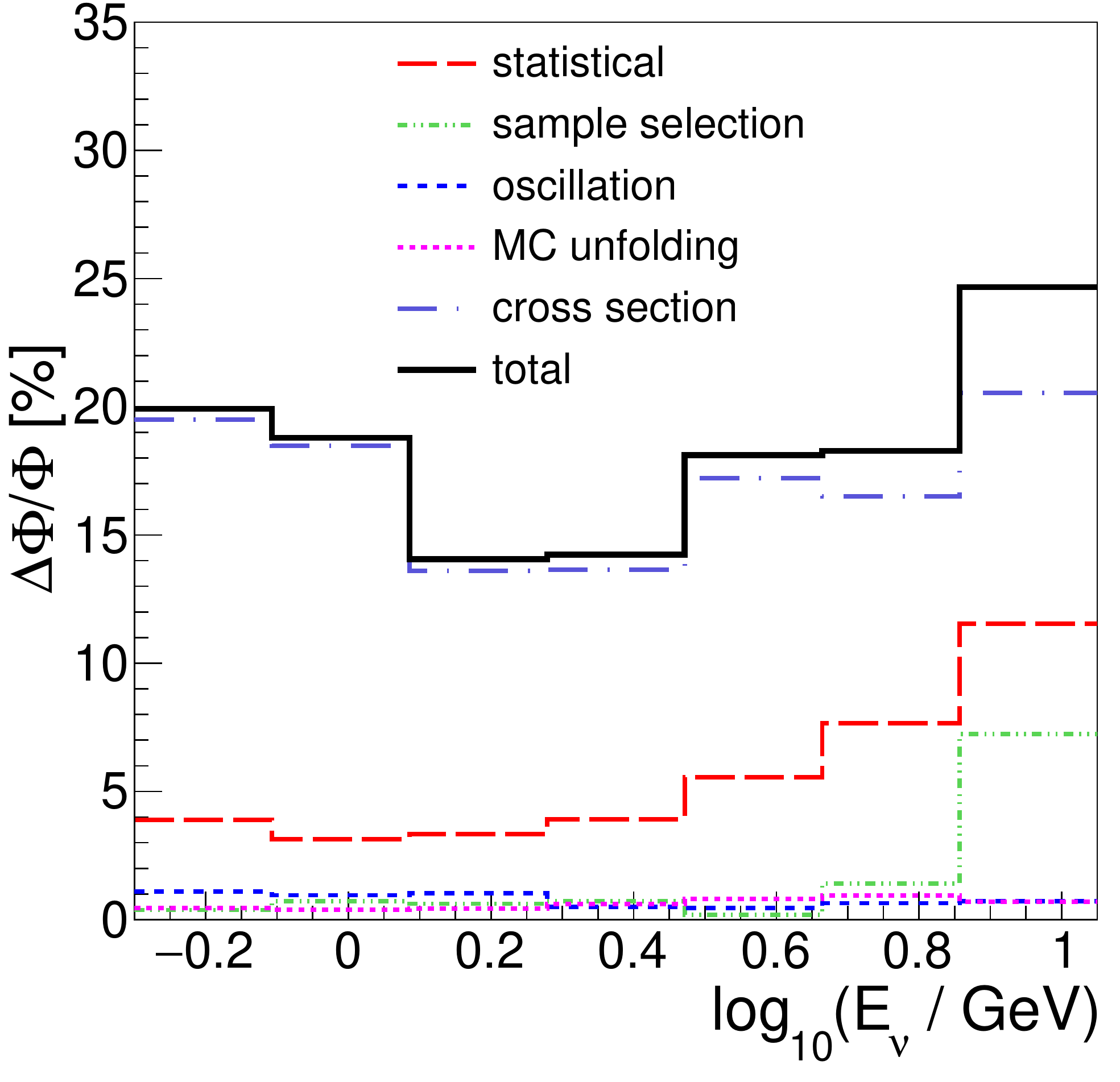}
\caption{Summary of estimated relative uncertainty on the unfolded flux, reported separately for each source. The total relative uncertainty is also reported. Left: $\nu_e$ spectrum; right: $\nu_\mu$ spectrum.}
\label{fig:errsummary}
\end{center}
\end{figure}

 \indent The contributions of each uncertainty source are reported in Figure \ref{fig:errsummary}, for each unfolding bin. The total uncertainty reported is calculated as the sum in quadrature of all contributions. The neutrino cross-section uncertainty represents the dominant contribution over the whole unfolded spectrum. The statistical uncertainty has also an important weight in high-energy bins. The total flux uncertainty ranges from a minimum value of 10-15\% in the $\mathcal{O}$(1\,GeV) energy region, up to a 20-25\% in the edge bins.

\section{Results and discussion}
\begin{figure}[h!]
    \centering
    \includegraphics[width = 0.495\textwidth]{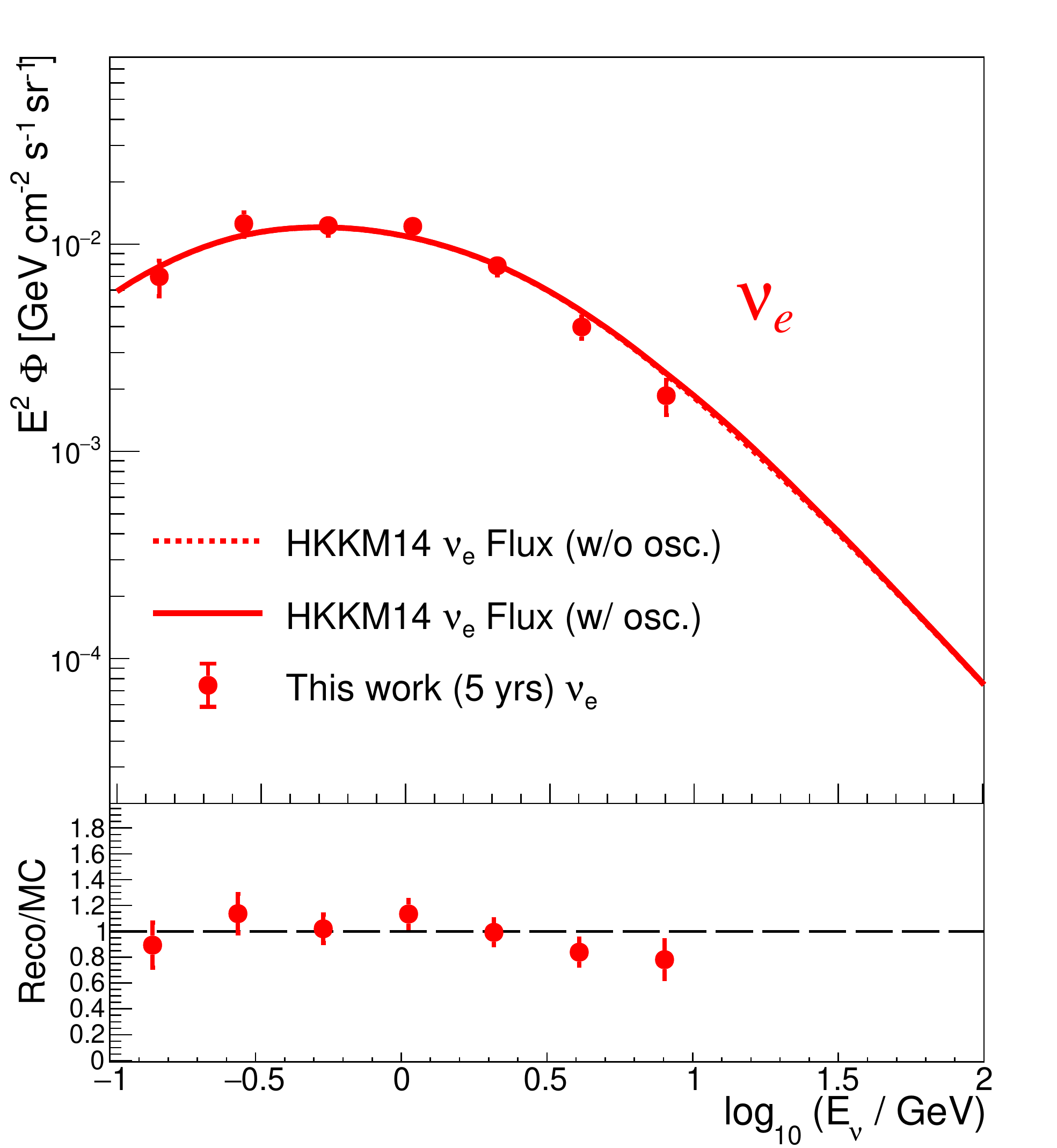}
    \includegraphics[width = 0.495\textwidth]{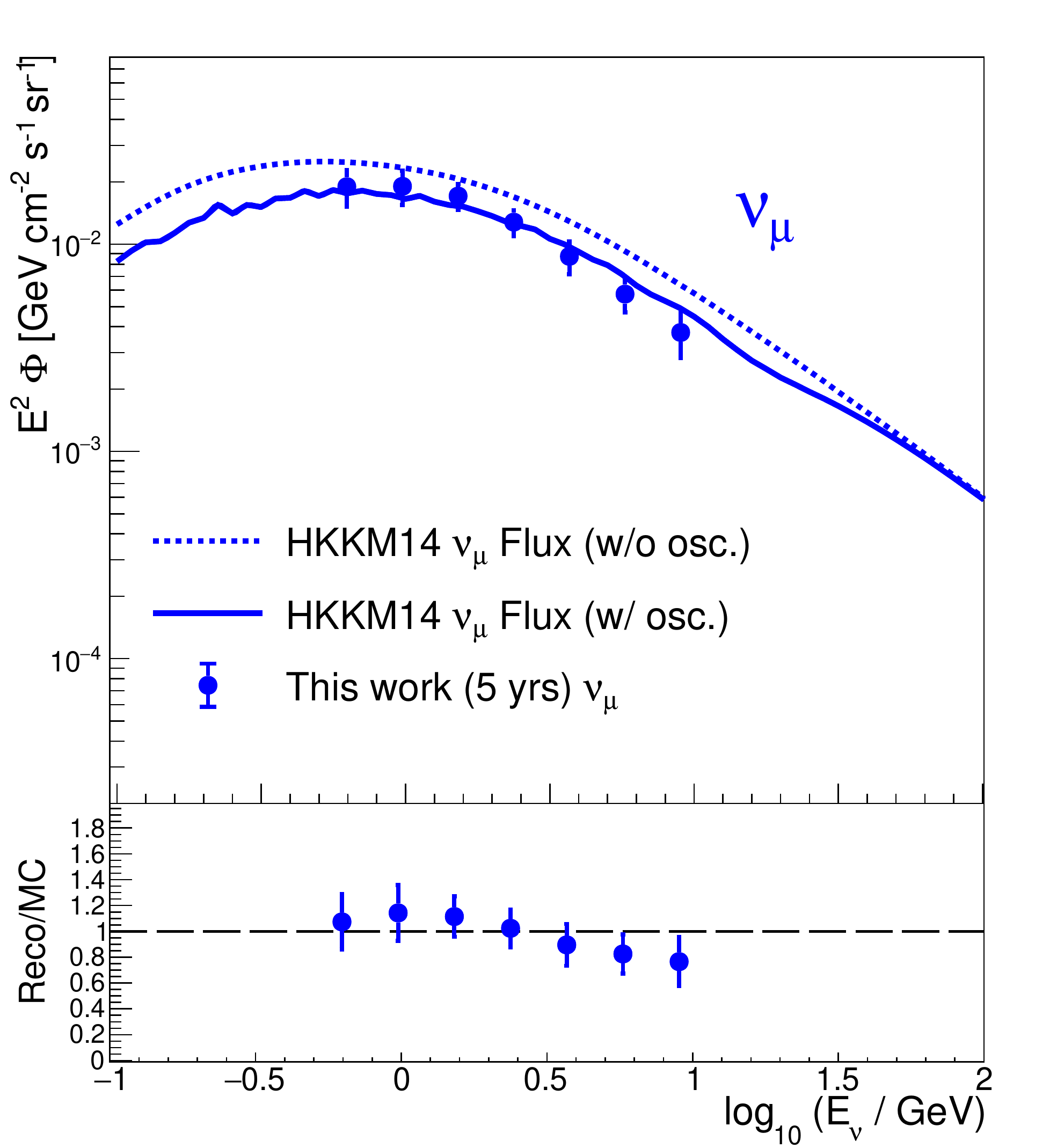}
    \caption{Reconstructed energy spectra for the $\nu_e$ (left) and the $\nu_\mu$ flux (right). The fluxes are plotted multiplied by $E^2$, to give a better picture. The error bars on the flux values include all statistical and systematic contributions evaluated in Section \ref{sect:err}. The HKKM14 flux prediction~\cite{HKKM14} is also reported, at the source (dashed line) and including the oscillation effects (full line).}
    \label{fig:nuspec}
\end{figure}

The unfolded \nel~and \nm~energy spectra are shown
In Figure \ref{fig:nuspec}. The binning is described in Section \ref{sect:unfolding}. The predicted HKKM14 flux~\cite{HKKM14} is also reported, both at the source and including oscillation effects along the baseline. The oscillation-induced flux deficit in the $\nu_\mu$ flux below 10\,GeV is clearly visible.

JUNO is able to reconstruct the energy spectrum of atmospheric neutrinos in the energy range [100\,MeV - 10\,GeV], usually referred to as the $``$low-energy'' region. This work, although based on simulated data only, shows the good capabilities of a large LS based detector like JUNO to measure the atmospheric neutrino flux. The energy region considered is already populated by other measurements, however some discrepancies still remain. JUNO can provide additional information about this interesting energy region, helping models in constraining their predictions.
The quoted uncertainty is competitive with present experimental results and shows a margin of improvement by the increase of exposure time. Although JUNO's design is not optimized for atmospheric neutrino physics, 
the extremely good performances in the atmospheric neutrino energy reconstruction can be fully exploited for the measurement of the energy spectrum.
Moreover, atmospheric neutrinos are a natural source which will be fully accessible from the beginning of data taking.

\section{Conclusions}

\noindent The JUNO detector has been designed from the beginning as a 
state-of-the-art detector for neutrino physics. The large dimensions of the 
detector, as well as its dense instrumentation, pave the way to an entire 
series of measurements, in a multi-purpose approach. The atmospheric 
neutrino flux is a natural source that can be observed, from the very 
beginning of data-taking. Although the detector design is not optimized for 
this class of events, the large active volume and the fine energy resolution
allow to reconstruct the energy spectrum with a competitive precision, 
especially in the low-energy region.

In this work, a large set of MC events has been generated to evaluate the 
detector performances. A smaller set has been used to simulate the real 
data. Thanks to the timing performances of JUNO, the flavor of primary 
neutrinos can be separated with a limited residual contamination. 
A rejection power of the order of $10^{5}$ has been applied to reduce 
atmospheric muon background.
The atmospheric neutrino energy spectrum has been reconstructed in the 
energy range [100\,MeV - 10\,GeV], separately for $\nu_e$ and $\nu_\mu$, 
assuming a $\sim$5 years detector livetime. The reconstructed spectra lie 
inside an interesting energy region, where previous measurements show some 
discrepancies. The results obtained show the good performance of JUNO in detecting the atmospheric neutrino flux in the low energy region, where 
theoretical models have large uncertainties. The inferred information can 
provide a fruitful input to constrain flux predictions, which are essential 
to evaluate the impact of atmospheric neutrinos in the search of rare 
events. 

\section*{Acknowledgements}
We are grateful for the ongoing cooperation from the China General Nuclear Power Group. This work was supported by the Chinese Academy of Sciences, the National Key R\&D Program of China, the CAS Center for Excellence in Particle Physics, Wuyi University, and the Tsung-Dao Lee Institute of Shanghai Jiao Tong University
in China, the Institut National de Physique Nucléaire et de Physique
de Particules (IN2P3) in France, the Istituto Nazionale di Fisica Nucleare (INFN) in Italy, the Italian-Chinese collaborative research program
MAECI-NSFC, the Fond de la Recherche Scientifique (F.R.S-FNRS)
and FWO under the “Excellence of Science - EOS in Belgium, the
Conselho Nacional de Desenvolvimento Científico e Tecnològico in
Brazil, the Agencia Nacional de Investigacion y Desarrollo in Chile,
the Charles University Research Centre and the Ministry of Education,
Youth, and Sports in Czech Republic, the Deutsche Forschungsgemeinschaft (DFG), the Helmholtz Association, and the Cluster of Excellence
PRISMA+ in Germany, the Joint Institute of Nuclear Research (JINR)
and Lomonosov Moscow State University in Russia, the joint Russian
Science Foundation (RSF) and National Natural Science Foundation of
China (NSFC) research program, the MOST and MOE in Taiwan, the
Chulalongkorn University and Suranaree University of Technology in
Thailand, and the University of California at Irvine in USA.

\bibliography{Atmo_nu_juno}

\onecolumn
\begin{flushleft}
{\Large The JUNO Collaboration}\\
\bigskip
Angel Abusleme$^{5}$, 
Thomas Adam$^{45}$, 
Shakeel Ahmad$^{66}$, 
Rizwan Ahmed$^{66}$, 
Sebastiano Aiello$^{55}$, 
Muhammad Akram$^{66}$, 
Fengpeng An$^{29}$, 
Guangpeng An$^{10}$, 
Qi An$^{22}$, 
Giuseppe Andronico$^{55}$
% ...
\\
\bigskip
% ...
$^{5}$Pontificia Universidad Cat\'{o}lica de Chile, Santiago, Chile\\
% ...
$^{10}$Institute of High Energy Physics, Beijing, China\\
% ...
$^{22}$University of Science and Technology of China, Hefei, China\\
% ...
$^{29}$East China University of Science and Technology, Shanghai, China\\
% ...
$^{45}$IPHC, Universit\'{e} de Strasbourg, CNRS/IN2P3, F-67037 Strasbourg, France\\
% ...
$^{55}$INFN Catania and Dipartimento di Fisica e Astronomia dell Universit\`{a} di Catania, Catania, Italy\\
% ...
$^{66}$Pakistan Institute of Nuclear Science and Technology, Islamabad, Pakistan\\
% ...

\end{flushleft}

\end{document}